\documentclass[a4paper,aps,pre,floatfix,twocolumn,showpacs]{revtex4}

\usepackage[dvips]{epsfig}
\usepackage[dvips]{graphicx}
\usepackage{latexsym,amsmath,verbatim}
\usepackage[dvips,usenames]{color}

\begin{document}

\title{Role of feedback and broadcasting in the naming game}

\author{Andrea Baronchelli} 
\affiliation{Departament de F\'\i sica i Enginyeria Nuclear, Universitat
  Polit\`ecnica de Catalunya, Campus Nord B4, 08034 Barcelona, Spain}

\begin{abstract}
\noindent The naming game (NG) describes the agreement dynamics of a population of agents that interact locally in a pairwise fashion, and in recent years statistical physics tools and techniques have greatly contributed to shed light on its rich phenomenology.  Here we investigate in details the role played by the way in which the two agents update their states after an interaction. We show that slightly modifying the NG rules in terms of which agent performs the update in given circumstances (i.e. after a success) can either alter dramatically the overall dynamics or leave it qualitatively unchanged. We understand analytically the first case by casting the model in the broader framework of a generalized NG. As for the second case, on the other hand, we note that the modified rule reproducing the main features of the usual NG corresponds in fact to a simplification of it consisting in the elimination of feedback between the agents. This allows us to introduce and study a very natural broadcasting scheme on networks that can be potentially relevant for different applications, such as the design and implementation of autonomous sensor networks, as pointed out in the recent literature.

\end{abstract}


%64.60.Cn      %Order-disorder transitions, 64.60.Cn
%89.75.-k        %Complex systems, 89.75.-k
%05.65.+b       %statistical physics, 05.65.+b
%89.75.Hc 	%Networks and genealogical trees 
%89.65.-s        %social systems
%05.65.+b	%Self-organized systems
%89.75.-k	       %Complex systems (for complex chemical systems, see 82.40.Qt; for biological complexity, see 87.18.-h)
%89.75.Da	%Systems obeying scaling laws
%89.75.Fb	%Structures and organization in complex systems
%89.75.Hc	%Networks and genealogical trees
%89.75.Kd	%Patterns
%64.70.Q- 	%Theory and modeling of the glass transition 
%05.40.Fb 	%Random walks and Levy flights 
%64.60.aq 	%Networks
%89.75.-k 	%Complex systems
%89.75.Hc 	%Networks and genealogical trees 
\maketitle

\section{Introduction}

The naming game (NG)~\cite{Steels1996,Baronchelli_JStatMech_2006} 
describes a population of agents playing pairwise interactions in order to 
{\em negotiate} conventions. Following Wittengstein's intuition on language \cite{wittgenstein53english},
the negotiation is seen as an activity in which one of the individuals 
(i.e. the ``speaker'') tries to draw the attention of the other (i.e. the ``hearer'') 
towards an external meaning through the production of a conventional 
form. For example, the speaker might want to make the hearer identify an object 
trough the production of a name. Based on the success or failure of the hearer
in pinpointing the proper meaning, both agents reshape their internal meaning-form
associations. Since in general at each time step a different pair interacts,
the interesting point is to study how the local dynamics affects
the population scale behavior, and to investigate the mechanisms leading to 
the final global consensus. 

The model, originally defined to describe artificial intelligence experiments \cite{Steels1996}, has been recently brought 
to the attention of the community of statistical physicists~\cite{Baronchelli_JStatMech_2006} 
in a formulation that is very close in spirit to that of other opinion dynamics models~\cite{fu2008coevolutionary,blythe2009gmc}
(for a detailed analysis of this point see  \cite{Castellano_2009}). It has been studied in fully 
connected graphs (i.e. in mean-field or homogeneous mixing populations) \cite{Steels1996,Baronchelli_JStatMech_2006,Baronchelli_ng_long}, 
regular lattices \cite{Baronchelli_2005,lu2008naming}, small world networks \cite{dallasta06,lu2008naming,liu2009naming}, random geometric graphs \cite{lu2006naming,lu2008naming,jiabo2010} and static \cite{dallasta06b,dallasta06c,yang2008}, dynamic \cite{nardini2008s} and empirical \cite{lu2009naming} complex networks.  The final state of the system is always 
consensus~\cite{de2006reach}, but stable polarized states can be reached introducing 
a simple confidence/trust parameter \cite{Baronchelli_2007}. The naming game as defined in~\cite{Baronchelli_JStatMech_2006} has also been modified in several ways~\cite{lu2006naming,Baronchelli_2007,wang2007,brigatti2008consequence,lipowski2008bio,lipowski2009,lu2008naming,brigatti2009conventions,lu2009naming,Lei20104046,zhang2010noise,yong2010,lei_ng_weights} and it represents the fundamental brick of more complex models in computational cognitive sciences \cite{puglisi08,baronchelli10}.	 From the point of view of the applications,
finally, it has been pointed out its relevance in system-design in the context of sensor networks \cite{akyildiz2002survey}, for such problems as autonomous key creation or selection for encrypted communication  \cite{lu2008naming} and, more recently, as a tool to investigate the community structure of social networks \cite{lu2009naming,zhang2010noise}.

\begin{figure}[t]
%\begin{center}
\includegraphics*[width=0.42\textwidth]{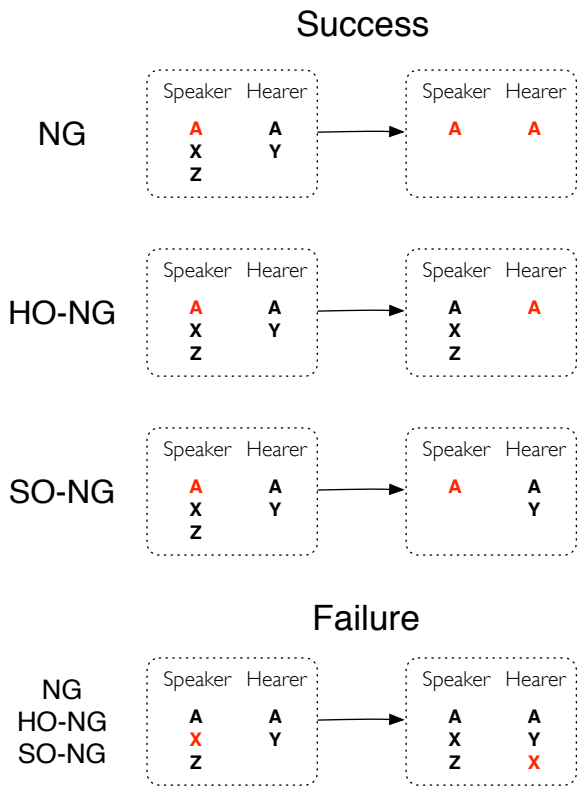}
%\end{center}
\caption{(Color online) \textbf{Interaction rules.} The speaker's inventory contains three words, namely $A$, $X$ and $Z$, while the hearer's one contains $A$ and $Y$. If the speaker randomly selects word $A$, the interaction is a success (Top). In the usual NG both agents delete competing synonyms, in the HO-NG only the hearer update her inventory, while in the SO-NG only the speaker does that. When the word transmitted by the speaker is not known by the hearer (as for example $X$), on the other hand, the latter adds it in her inventory and the interaction is classified as a failure (Bottom). In this case, both the HO-NG and the SO-NG behave as the NG.}
 \label{f:rules}
 \end{figure}

The rules are simple~\cite{Baronchelli_JStatMech_2006}. The game is played by a population of $N$ agents, each of which characterized by an inventory, i.e. a list of words (or ``conventions", ``opinions'', ``forms" or ``states"), whose size is not fixed. At every time step two agents are randomly selected and interact (see also Figure \ref{f:rules}). One of them plays as speaker and the other one as hearer.
The speaker picks randomly a word from her inventory and conveys it to the hearer. If the hearer's inventory contains that word, the game is a success, and both agents delete all the words in their inventories but the one that has just been transmitted. Otherwise it is a failure, and the hearer adds the received word to her inventory. The scheme is completed by specifying that at the beginning of the game all inventories are empty, and that whenever a speaker has an empty inventory she invents a brand new word and transmits it to the hearer \footnote{The fact that the invented word is actually new to the whole population corresponds to discarding homonymy~\cite{Baronchelli_JStatMech_2006}.}.

Here we focus on the success rule. The fact that both agents undergo the very same operation (i.e. shrink their inventories to the same unique word), underlies the existence of a feedback between the two. In the original formulation the feedback occurrs through an outside world, with the hearer pointing to the object she would associate with the received word. The speaker would then point on its turn to the right object and both individuals would immediately know whether the game was a success or a failure \cite{Steels1996}. In the simplified version defined in~\cite{Baronchelli_JStatMech_2006}, however, the feedback simply consists in the hearer informing the speaker that she too has the transmitted word. In case of failure, on the other hand, no feedback is needed. 

In what follows, we will investigate what happens when only one of the agents updates her inventory after a successful interaction. We will show that the situation changes dramatically depending on whether the update is performed by the hearer only or the speaker only, cases to which we will refer to as Hearer Only NG (HO-NG) and Speaker-Only NG (SO-NG) respectively (see Figure~\ref{f:rules}). In particular, we will show that the HO-NG yields a scaling of the convergence time with the population size that is the same as the one observed in the usual NG. The SO-NG, on the other hand is significantly slower. We will understand analytically the reason beyond this difference and point out that the SO-NG spontaneously falls in the critical regime of the generalized naming game model introduced in \cite{Baronchelli_2007} (Sec. II). The fact that the HO-NG remains efficient, on the other hand, will allow us to introduce a very natural broadcasting scheme that significantly simplifies previously introduced protocols \cite{lu2008naming} when the population is embedded in any kind of topology (Sec III). 

\section{Non-symmetric updating and the role of feedback}

\begin{figure}[t]
%\begin{center}
\includegraphics*[width=0.45\textwidth]{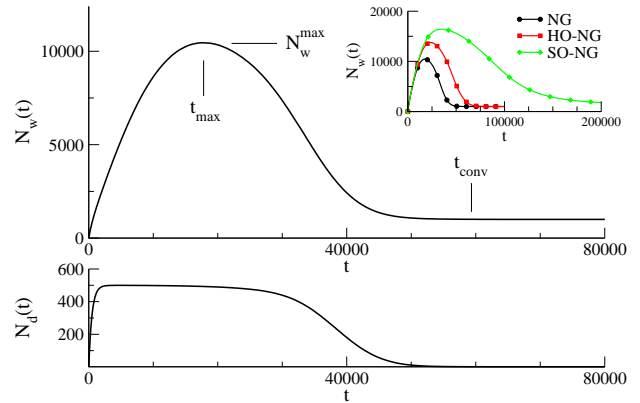}
%\end{center}
\caption{(Color online) {\bf Time evolution of the naming game.} Top: The total number of words $N_w(t)$ grows till it reaches a maximum, $N_w^{max}$  at time $t_{max}$, corresponding to the maximum amount of memory required to the population. Due to an increase in successful interactions, it then start decreasing till the final state in which $N_w(t)=N$, meaning that all the agents share the same unique word. The Inset shows that the HO-NG and the SO-NG exhibit the same qualitative behavior observed in the NG. Bottom: the number of different words present in the system saturates at $N_d(t)=N/2$ as soon as the invention process ceases. It then remains almost constant to fall finally to the consensus value $N_d(t)=1$ at the convergence time $t_{conv}$. 
Data refer to a population of $N=10^3$ agents and are averaged over $10^3$ realizations. }
 \label{f:curve}
 \end{figure}

Relevant observables in the NG are the total number of words $N_w(t)$, defined as the sum of the inventory sizes of all the agents, and the number of different words $N_d(t)$, counting, as the name suggests, how many different words are present in the system at time $t$~\cite{Baronchelli_JStatMech_2006}. 
%Thus, in a population of $N=2$ agents whose inventories of size one containthe same word it would be $N_w(t)=2$ and $N_d(t)=1$. 
The dynamics proceeds as follows (see Figure~\ref{f:curve})~\cite{Baronchelli_JStatMech_2006,Baronchelli_ng_long}: at the beginning both  $N_w(t)$ and $N_d(t)$ grow linearly as the agents invent new words. As invention ceases, $N_d(t)$ reaches a plateau whose height is on average $N/2$ words (since the agents interact in pairs). $N_w(t)$ keeps on the other hand growing till it reaches a maximum at time $t_{max}$, whose height, $N_w^{max}$, corresponds to the highest amount of memory (i.e. size of the inventories) required to the system.
The total number of words then decreases and the system reaches the convergence state at time $t_{conv}$. At convergence all the agents share the same unique word, so that $N_w(t_{conv})=N$ and $N_d(t_{conv})=1$. 

\begin{figure}[t]
%\begin{center}
\includegraphics*[width=0.45\textwidth]{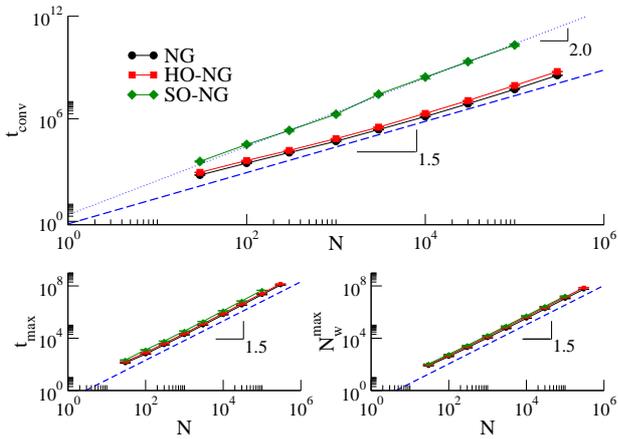}
%\end{center}
\caption{(Color online) {\bf Scaling with the system size.} Top: In the SO-NG the convergence time $t_{conv}$ is much larger than in the NG and the HO-NG. Also the scaling with the population size is slower than the $t_{conv} \sim N^{\omega}$ with $\omega \simeq 1.5$ (dashed line) exhibited by both the NG and the HO-NG. Bottom panels: the time (left) and the height (right) of the total number of words, on the other hand, scale in all cases as $t_{max} \sim N^{\nu}$ and $N_w^{max} \sim N^{\mu}$ with $\nu \simeq \mu \simeq 1.5$ (dashed lines). Each point represents the average value obtained over $30$ simulation runs and is plotted with the correspondent statistical error (often not visible on the scale of the graph).}
 \label{f:scaling_asym}
 \end{figure}

We now focus on the scaling of the memory consumption and the convergence time with the system size. In the usual NG it holds $t_{conv} \sim N^\omega$, $t_{max} \sim N^\nu$ and $N_w^{max} \sim N^\mu$, with $\omega \simeq \mu \simeq \nu \simeq 1.5$ (where the time is counted in terms of microscopic interactions) \cite{Baronchelli_JStatMech_2006}. Figure \ref{f:scaling_asym} shows that the same scaling is observed also in the SO-NG and the HO-NG as far as the time and height of the peak of $N_w(t)$ are concerned. Looking at the convergence time, however, it is clear that while the HO-NG behaves as the usual NG \footnote{The curve for the convergence time of the NG for systems of different sizes exhibits small oscillations when plotted in log-log scale \cite{Baronchelli_ng_long} and it does not therefore appear as a perfectly straight line.}, the SO-NG is remarkably slower, showing a $t_{conv} \sim N^\omega$ with $\omega \simeq 2.0$ behavior.
%(as observed for instance in the voter model \cite{Holley1975,cox1989coalescing,Castellano_2009}).
This numerical result is important. Indeed, the fact that that the HO-NG behaves substantially in the same way as the usual NG implies that the hearer's feedback to the speaker is not crucial, and opens the way to the implementation of straightforward broadcasting protocols, as we will see in the next section \footnote{The fact that the HO-NG and the NG behave in the same way as far as the scaling with the system size of the relevant quantities is concerned holds in the framework of the simplified scheme introduced in~\cite{Baronchelli_JStatMech_2006}, being a consequence of the the fact that homonymy is not taken into account. Feedback remains in fact a fundamental ingredient of any language games, as devised by Wittengstein \cite{wittgenstein53english}.}.
 
To shed light on the extremely slow convergence of the SO-NG it is convenient to consider the generalized NG scheme defined in \cite{Baronchelli_2007}. The rules are the same as in the NG, but for the fact that in case of a successful interaction the agents update their inventories with probability $\beta$ (so that the usual rules correspond to the $\beta=1$ case).
Generalizing in the same way the HO-NG and the SO-NG is straightforward: in the first, only the hearer will update her inventory after a success and will do that with a probability $\beta$, in the latter it will be only the speaker. The generalized NG exhibits a consensus-polarization transition at $\beta_c=1/3$. For $\beta>\beta_c$ the system always reaches the final consensus state, while for $\beta<\beta_c$ two competing words survive asymptotically (in the limit $N \rightarrow \infty$) and convergence is never reached \footnote{This is in fact the first of a series of transitions occurring as $\beta \rightarrow 0$  \cite{Baronchelli_2007}.}. The transition can be understood considering that, after the peak of the total number of words, the dynamics proceeds through the progressive elimination of competing words, and just before convergence only two different words, say $A$ and $B$, are present in the system~\cite{Baronchelli_ng_long}. Thus, the population can be divided into three groups formed by all the agents whose inventory stores either only $A$ or only $B$ or both $A$ and $B$, whose relative size in the population we label as $n_A$, $n_B$ and $n_{AB}$. The transition probabilities from different groups are the following \cite{Baronchelli_2007,Castello:2009p6}:

\begin{eqnarray}
 p_{A\to AB} = n_{B} + \frac{1}{2} n_{AB},  \;\; p_{B\to AB} =
  n_A + \frac{1}{2} n_{AB} \label{eq:NGa} \;\;\;\; \\
  p_{AB \to A} = \frac{3 \beta}{2} n_A + \beta n_{AB}, \;\; p_{AB \to B} = \frac{3 \beta}{2} n_B + \beta n_{AB} \label{eq:NGb}
\end{eqnarray}

\noindent In equation~(\ref{eq:NGa}), an agent with a single word $A$
($B$) adds $B$ ($A$) to its inventory when, playing as hearer, she
receives it from the speaker. This may happen either because the
speaker stores only $B$ ($A$), or because the speaker stores both
words and selects randomly, with probability $1/2$, $B$ ($A$).
In equation~(\ref{eq:NGb}), on the other hand, an agent reduces her
inventory from $AB$ to $A$ ($B$) only in case of an update
following, with probability $\beta$, a success on word $A$
($B$). The interaction may involve another agent with two
words (in this case the factor $1/2$ relative to the speaker
extraction is balanced by the fact that both agents will have only
$A$ ($B$) in their inventory), or an agent storing only $A$
($B$). In the latter case, either the $AB$ agent is the speaker and
plays $A$ ($B$) with probability $1/2$ or she is the hearer and the
success is certain, the sum of these two terms yielding the factor
$3/2$ (equation~(\ref{eq:NGb})).

The above transition probabilities translate into the following mean-field equations for the evolution in time of the fractions of agents in each state \cite{Baronchelli_2007}:

\begin{eqnarray}
\frac{d n_A}{dt} = - n_A n_B + \beta n_{AB}^{2} + \frac{3 \beta -1}{2}n_A n_{AB} 
\label{eq:meanfield_NG_a} \\
\frac{d n_B}{dt} = - n_A n_B + \beta n_{AB}^{2} + \frac{3 \beta -1}{2}n_B n_{AB}
\label{eq:meanfield_NG_b}
\end{eqnarray}

\noindent and $n_{AB}=1-n_A-n_B$. The fixed points are $n_A=1, n_B=n_{AB}=0$ and $n_A=n_{AB}=0, n_B=1$ and $n_A=b(\beta),n_B=b(\beta),n_{AB}=1-2b(\beta)$ with
$b(\beta)=\frac{1+5\beta-\sqrt{1+10\beta+17\beta^2}}{4\beta}$ (and
$b(0)=0$). Defining the magnetization as $m=n_A-n_B$, we have

\begin{eqnarray}
  \frac{d m}{dt} = \frac{3\beta-1}{2}n_{AB}m.
  \label{eq:magnetization_NG} 
\end{eqnarray}

\noindent Thus, for
$\beta_c>1/3$, $sign(\frac{d m}{dt})=sign(m)$: 
$|m|\rightarrow 1$ and the system ends up in  an absorbing state of
consensus in the A or B option. For $\beta_c<1/3$, $sign(\frac{d
  m}{dt})=-sign(m)$ and $|m|\rightarrow 0$, implying the stationary
coexistence of the three phases, with $n_A=n_B$ and a finite density
of AB agents.

\begin{figure}[t]
%\begin{center}
\includegraphics*[width=0.42\textwidth]{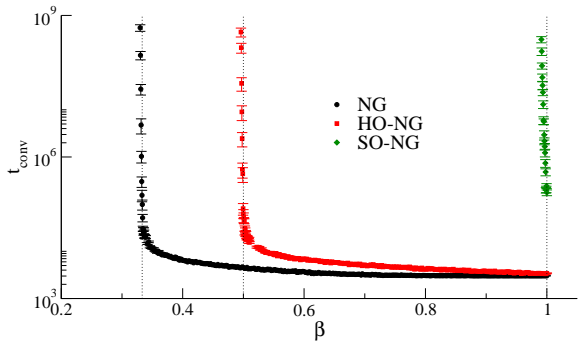}
%\end{center}
\caption{(Color online) {\bf The generalized NG - Consensus time as a function of $\beta$.} In the NG the convergence time $t_{conv}$ diverges at $\beta_c=1/3$. The HO-NG is less robust, and $\beta_c=1/2$. For the SO-NG, on the other hand, it holds $\beta_c=1$, so that the final consensus is never a stable solution. Dotted vertical lines represent the theoretical predictions. Data refer to a population of $3 \times 10^3$ agents and each point represents the average value obtained from $30$ simulation runs. Error bars represent the statistical error of the average value.}
 \label{fig:diver_beta}
 \end{figure}
 
To describe the SO-NG in the same framework one must modify Eq.~(\ref{eq:NGb})
to take into account that the hearer never updates her inventory after a success [Eq.~(\ref{eq:NGa}), concerning the update following a failure, remains unchanged]. The transition probabilities now read 

\begin{eqnarray}
  p_{AB \to A} = \beta \left(\frac{n_A}{2} + \frac{n_{AB}}{2} \right), \; p_{AB \to B} = \beta \left(\frac{n_B}{2} +  \frac{n_{AB}}{2} \right)
\end{eqnarray}

\noindent yielding 

\begin{eqnarray}
\frac{d n_A}{dt} = - n_A n_B + \frac{\beta}{2} n_{AB}^{2} + \frac{ \beta -1}{2}n_A n_{AB} 
\label{eq:meanfield_NG_a} \\
\frac{d n_B}{dt} = - n_A n_B + \frac{\beta}{2} n_{AB}^{2} + \frac{ \beta -1}{2}n_B n_{AB}
\label{eq:meanfield_NG_b}
\end{eqnarray}

\noindent and therefore 

\begin{eqnarray}
  \frac{d m}{dt} = \frac{\beta-1}{2}n_{AB}m.
  \label{eq:magnetization_NG} 
\end{eqnarray}

\noindent Thus, the transition occurs at $\beta_c=1$, and the SO-NG is naturally critical. This explains why the behavior of the SO-NG is qualitatively different from the one observed in the usual NG. The system is not driven to consensus, but rather ends up there only due to large, system size, fluctuations of the magnetization. The same analysis can be repeated also for the generalized HO-NG, where one finds that $\beta_c=1/2$. Hence, even though for $\beta=1$ the HO-NG and the NG behave in the same qualitative way, the latter is more robust to perturbations in the generalized setting. Figure \ref{fig:diver_beta} shows that numerical simulations are in very good agreement with the theoretical prediction, confirming that focusing on the two words case is indeed a valid assumption to determine the critical values of $\beta$ (while it can be an oversimplifying starting point to describe more subtle properties of the convergence process in the NG \cite{Baronchelli_2007}).

\section{Broadcasting on networks}

\begin{figure}[t]
%\begin{center}
\includegraphics*[width=0.42\textwidth]{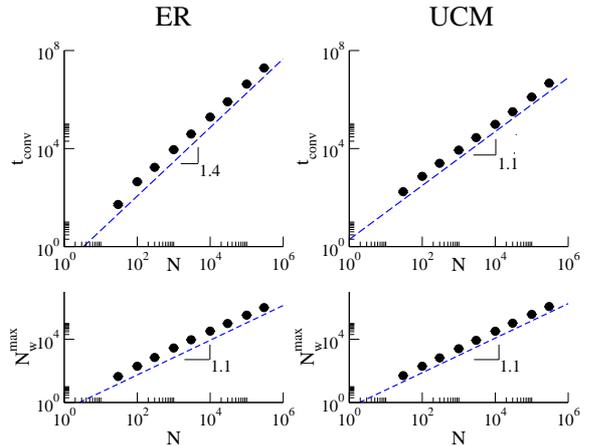}
%\end{center}
\caption{(Color online) {\bf Broadcasting on networks.} Scaling of the consensus time (top) and the maximum number of words (bottom) for the broadcasting rule on ER homogeneous (left) and UCM heterogeneous networks (right). Each point is averaged over $30$ simulation runs ($5$ runs on each of $6$ network realizations), and error bars are not visible on the scale of the graph. Heterogeneous networks (right panels) are generated with the Uncorrelated Configuration Model (UCM) \cite{ucmmodel}, with minimum degree $k=4$ and degree exponent, $P(k) \sim k^{- \gamma}$, is $\gamma=2.5$.  ER networks are generated with $\langle k \rangle = 8$.}
 \label{fig:broad_scaling}
 \end{figure}
 
Complex networks are the natural environment to investigate the dynamics of models that aim at describing social, techological or biological systems \cite{caldarelli2007sfn,barratbook,mendesbook,romuvespibook,boccaletti2006cns}.
Motivated by the communication protocols employed by sensor nodes, Lu, Korniss and Szymanski introduced a broadcast version of the NG to make it applicable in sensor networks \cite{lu2006naming,lu2008naming}. In this framework, the rules are the same as in the  usual NG but for the fact that the speaker transmits her word to all her neighbors at the same time, rather than to a randomly selected one. If a hearer has that word in her inventory, she deletes all the competing synonyms but the winning one, otherwise she adds the new word to her memory. As for the speaker, she will consider the interaction a success if at least one of the hearers knows it. At least one successful hearer must therefore report to the speaker that she knew the transmitted word. In real network this communication could be performed for example through the ``lecture hall" algorithm \cite{lecturehall_alg}.

The results discussed above about the HO-NG however show that, at least as far as pairwise interactions are considered, there is no need for the speaker to receive any feedback in order to guarantee an efficient route to convergence. Thus, a very natural broadcasting scheme might simply consist in letting only the hearer update their inventories, following the usual rules in case of failure or success. Figure \ref{fig:broad_scaling} shows that such a simple broadcasting scheme guarantees a fast convergence. The scaling exponent $\omega$ of the consensus time, $t_{conv} \sim N^{\omega}$, is $\omega \simeq 1.4$ in Erdos-Renyi (ER) homogeneous graphs and $\omega \simeq 1.1$ in uncorrelated heterogeneous networks \cite{ucmmodel}, the latter being compatible also with a logarithmic behavior $t_{conv} \sim N \ln N$. Therefore the broadcasting scheme yields a much faster convergence when the broadcasting protocol is adopted, but only in heterogeneous networks. This is not the case in the usual NG, where it holds $\omega \sim 1.4$ in both ER and heterogeneous networks \cite{dallasta06b}. 

Though faster, however, the broadcasting scheme is not scalable in the thermodynamic limit, since the maximum memory required to the system scales with an exponent $\mu \simeq 1.1$ (again compatible with a behavior $N_W^{max} \sim N \ln N$), thus implying that single agents should have an infinite memory as $N\rightarrow \infty$. This is not the case for the usual NG, where it holds $\mu \simeq 1$  \cite{dallasta06b}.
Strictly speaking, therefore, it is not possible to conclude whether the broadcast rule has to be preferred to the usual pairwise interaction scheme, but rather it would be necessary to decide in a case by case setting, depending also on the topology in which the agents are embedded. In case of heterogeneous networks,  broadcasting is certainly the fastest solution when the memory of the system is not a vital parameter, while pairwise interaction offers the best solution if inventory size is a (major) concern. However, the memory consumption is probably not a big issue for all the practical purposes in which a finite population is considered, due to the small value of $\mu$.

\begin{figure}[t]
%\begin{center}
\includegraphics*[width=0.42\textwidth]{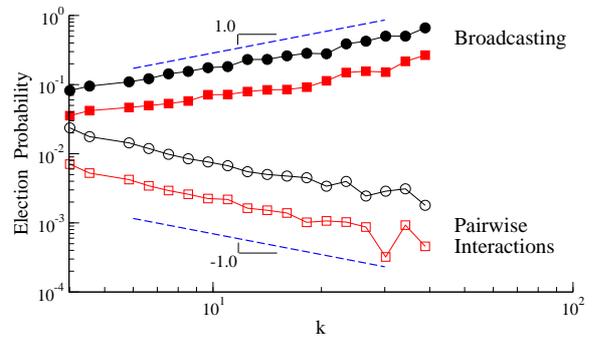}
%\end{center}
\caption{(Color online) {\bf Leader election.} The probability that a given agent gets elected increases (almost) linearly with the degree in the broadcasting scheme (full symbols) and it is inversely proportional to the same quantity in the usual NG with pairwise interactions (empty symbols). All curves refer to heterogeneous networks generated with the UCM with minimum degree $k=4$ and degree exponent $\gamma=2.5$ (circles) and $\gamma=3$ (squares). Data come from $10^4$ simulation runs and have been logarithmically binned. Curves are vertically shifted for clarity. }
 \label{fig:leader}
 \end{figure}

In \cite{lu2008naming} it is pointed out also that the NG could provide a valuable mechanism for \textit{leader election} among a group of sensors. The leader is a single node with important responsibilities ranging from routing coordination to key distribution, and the NG would make the identification of the leader hardly predictable from the outside, resulting in enhanced security of the system to possible attacks. To study the effect of broadcasting on the election of the leader, we have run simulations in which at the beginning of the process every node is endowed with a word. All words are different and therefore represent a tag assigned to the agents. Having checked that these initial conditions do not alter the scaling properties of the system (data not shown), we look at the statistics of the word upon which consensus is reached, and more in particular on the degree of the node to which it was assigned at the beginning. 
It has already been observed, even though never quantified, that in the case of pairwise interactions the hubs, playing mostly as hearers, are not good promoters for conventions, but rather act as intermediaries between lower degree nodes \cite{dallasta06b}. In fact, Figure \ref{fig:leader}  shows that in the NG the probability that a given node of degree $k$ spreads her identity (i.e., is elected) is linearly suppressed by a factor $k$, and the same behavior is observed also for the HO-NG (not shown). Remarkably, the situation is inverted in the broadcasting scheme, where this probability is magnified by the same linear $k$ factor. Due to the dramatic nonlinearities of the process it is difficult to go beyond the numerical experiment, but this observation can be relevant for the applications, and at the same time sheds light on the existence of profound differences between the pairwise and the broadcasting schemes at a microscopic level, whose detailed investigation is left for future work.

\section{Discussion and Conclusions}

In this paper we have focused on the role of feedback in the NG. We have modified the rule that prescribes the symmetric update of the agents' inventories after a successful interaction, addressing the two circumstances in which only one of them 
acts in case of success. We have shown that if the updating agent is the speaker the NG protocol becomes inefficient and the final consensus state is reached in much longer times as compared to the scaling of the usual symmetric rule. 
Casting this finding in the broader context of a generalized NG scheme, we have shown analytically that preventing the update of the hearer sets the model to a critical point below which consensus is not the stable solution, and the system would persist forever (in the thermodynamic limit) in a polarized state in which two words survive indefinitely. At the critical point the magnetization is therefore conserved and consensus is reached through large fluctuations in a finite system, thus being slower than in the NG. When the update is performed only by the hearer, on the other hand, the scaling of the main quantities with the system size remains the same as in the usual NG. Thus, feedback is not crucial in the NG as defined in \cite{Baronchelli_JStatMech_2006}, i.e. the scaling of the convergence time and of the memory required to the system does not change if the hearer never informs the speaker of the outcome of the game. 
%The model becomes therefore closer in spirit to the voter model \cite{Clifford73,holley1975etw}. 

The results concerning the HO-NG have also allowed us to introduce a very natural broadcasting scheme in which the speaker transmits simultaneously the word to all of her neighbors, which update their inventories following the usual rules. We have shown that this scheme is efficient in terms of convergence time, outperforming the pairwise interaction rule in heterogeneous networks as far as the scaling of the convergence time is concerned, with some minor drawbacks from the point of view of the memory requirements. Finally, we have discussed how these findings could be relevant also from the point of view of possible applications in the field of sensor networks  \cite{lu2006naming,lu2008naming}. 

Our work sheds light on the dynamics of the NG, pointing out that the update of the hearer is a fundamental ingredient of the model contrarily to the feedback provided to the speaker, which turns out to be less crucial. It also opens the way to several  lines of investigation. From the theoretical point of view, a systematic study of the broadcasting dynamics on different kinds of networks as well as its generalization, and thorough characterization, in the probabilistic $\beta$ framework are potentially very interesting. Furthermore, while in this paper we have concentrated on the study of the system scale behaviors, performing in future a detailed analysis of the microscopic aspects of the dynamics, so far addressed only in \cite{dallasta06c}, could provide important insights into the broadcasting rule as well as into the difference between the NG, the SO-NG and the HO-NG. It is also worth noting that the result concerning the possibility of neglecting the hearer's feedback might help in the more ambitious exploration of the connection between the dynamics of the NG and the one of the voter model \cite{Clifford73,holley1975etw} or of other simple ordering dynamics schemes (see also \cite{Castellano_2009} on the challenges concerning this point) . 

From the point of view of the applications, finally, the broadcasting scheme is relevant for sensor nodes, as we have already mentioned. Moreover,  it could be the best solution also in those cases in which the communicating agents are not embedded in a static network, but on the contrary move in an unknown environment and need to communicate about their exploration \cite{steels2003evolving}. In general, broadcasting is  the fundamental communication mechanism in different frameworks ranging from scientists communicating through articles visible to their community to social tagging systems like Delicious \cite{delicious}, from bacterial quorum sensing \cite{Waters:2005p112} to social networks like Facebook \cite{facebook} or Twitter \cite{twitter}, and the NG could now constitute an helpful conceptual tool also in these cases (see also \cite{cattuto2007semiotic}).

 \textit{Acknowledgments.} The author is indebted to Alain Barrat, Luca Dall'Asta, Vittorio Loreto, Luc Steels and Romualdo Pastor-Satorras for helpful discussions, and acknowledges support from the Spanish Ministerio de
  Ciencia e Innovaci\'{o}n through the Juan de la Cierva program, as
  well as from project FIS2010-21781-C02-01 (Fondo Europeo de
  Desarrollo Regional) and from the Junta de Andaluc\'{i}a project
  P09-FQM4682.

%\bibliographystyle{apsrev}
%\bibliography{biblio}

\end{document}